\documentclass[12pt]{iopart}

\usepackage{dcolumn}
\usepackage{bm}
\usepackage{ifpdf}
\usepackage{hyperref}
\usepackage{dcolumn}
\usepackage{bm}
\usepackage[spanish,english]{babel}
\usepackage{amsfonts}
\usepackage{amssymb}
\usepackage{graphicx}
\usepackage[latin1]{inputenc}
\newcommand{\LL}{\mathcal{L}}

\newcommand{\Hcal}{\mathcal{H}}

\newcommand{\be}{\begin{equation}}
\newcommand{\en}{\end{equation}}
\newcommand{\bea}{\begin{eqnarray}}
\newcommand{\ena}{\end{eqnarray}}

\begin{document}

\title{Robustness of braneworld scenarios against tensorial perturbations}

\author{D. Bazeia$^1$, L. Losano$^1$, R. Menezes$^{2,3}$, Gonzalo J. Olmo$^{4,1}$, and D. Rubiera-Garcia$^5$}
\address{$^1$Departamento de F\'isica, Universidade Federal da
Para\'\i ba, 58051-900 Jo\~ao Pessoa, Para\'\i ba, Brazil}
\address{$^2$Departamento de Ci\^encias Exatas, Universidade Federal da
Para\'iba, 58297-000 Rio Tinto, PB, Brazil}
\address{$^3$Departamento de F\'\i sica, Universidade Federal de Campina
Grande, 58109-970 Campina Grande, PB, Brazil}
\address{$^4$Departamento de F\'{i}sica Te\'{o}rica and IFIC, Centro Mixto Universidad de Valencia - CSIC.
Universidad de Valencia, Burjassot-46100, Valencia, Spain}
\address{$^5$Center for Field Theory and Particle Physics and Department of Physics, Fudan University, 220 Handan Road, 200433 Shanghai, China}

\eads{\mailto{bazeia@fisica.ufpb.br}, \mailto{losano@fisica.ufpb.br}, \mailto{rmenezes@dce.ufpb.br}, \mailto{gonzalo.olmo@csic.es}, \mailto{drubiera@fudan.edu.cn}}

\begin{abstract}
Inspired by the peculiarities of the effective geometry of crystalline structures, we reconsider thick brane scenarios from a metric-affine perspective. We show that for a rather general family of theories of gravity, whose Lagrangian is an arbitrary function of the metric and the Ricci tensor, the background and scalar field equations can be written in first-order form, and tensorial perturbations have a non negative definite spectrum, which makes them stable under linear perturbations regardless of the form of the gravity Lagrangian. We find, in particular, that the tensorial zero modes are exactly the same as predicted by Einstein's theory regardless of the scalar field and gravitational Lagrangians. 
\end{abstract}

\pacs{04.40.Nr, 04.50.-h, 04.50.Kd, 04.70.Bw}

\submitto{\CQG}

\maketitle

\section{Introduction}

The number of spatial dimensions of our Universe is a key foundational question that has received enormous theoretical and experimental attention in the last decades \cite{ed}. Initially inspired by developments in string theories, the idea of braneworlds as four-dimensional hypersurfaces embedded in a higher-dimensional (typically $5-$dimensional) bulk space has been explored with exquisite detail in the literature \cite{rs,bw}. Among the resulting constructions we find particularly interesting the case of thick branes. This setup consists on a $5D$ bulk space with a scalar field localized in the fifth dimension and whose energy density is  typically concentrated around the point $y=0$ of the extra spatial dimension, where the 4D brane resides. The rest of the matter fields are confined on this (thick) hypersurface, with the scalar field somehow acting as the force responsible for their confinement. This provides an interesting framework, where both the fermion mass hierarchy and proton stability can be investigated due to the thickness of the brane; see, e.g., Ref.~\cite{ahs}.

The notion of a higher dimensional space-time is so fundamental that transcends its original motivations and justifies its exploration from different perspectives. We thus find it useful to seek for inspiration in a lower dimensional analogy borrowed from condensed matter systems and interpret braneworld models as a kind of sandwich configuration in a crystal, with a (possibly thick) $2-$dimensional distribution of impurities within a $3-$dimensional crystalline structure, being the crystal analogous to the bulk and the (thick) layer of impurities the brane. Interestingly, the structure of crystals in the continuum limit admits an effective geometric description \cite{Kroner, Kleinert,Classics}. This point is relevant in our discussion of braneworlds because the presence of defects such as interstitials (either impurities or not) or vacancies in crystals requires the use of metric-affine geometry for a consistent description of their effective geometry. Indeed, a direct correspondence between the metric-affine geometry of crystals with point-like defects (interstitials and/or vacancies) and theories of gravity with non-metricity has been recently established \cite{Lobo:2014nwa}. It turns out that the density of point-like defects in a crystal is analogous to the density of energy-momentum in the gravitational context, both being responsible for the existence of non-metricity, a property associated with the affine connection. The confinement of  energy-momentum at or around a region, defining in this way the brane, is thus analogous to the concentration of point-like defects on a sandwich layer in a crystal. This analogy, therefore, provides a novel motivation for the study of braneworlds in geometric scenarios with independent metric and affine degrees of freedom. Given that the question of whether the space-time geometry is Riemannian or otherwise must be determined by experiments, rather than by tradition or convention, we believe that the impact that metric-affine geometry could have on the phenomenology of higher-dimensional models of the Universe should be explored in some detail. This is the main motivation for this work.

The dynamical laws governing the higher-dimensional scenario we are about to consider are, in principle, unknown. Nevertheless, it is reasonable to expect them to be close to those of General Relativity (GR). For this reason, and to allow for some degree of generality, we may assume that the gravity Lagrangian is some function of the space-time metric and the Ricci tensor, $F(g_{\alpha\beta},R_{\alpha\beta})$, being GR the case $F(g_{\alpha\beta},R_{\alpha\beta})=g^{\alpha\beta}R_{\alpha\beta}$. We will see that, unlike in the (more standard) metric formulation of such theories, the field equations for this type of theories are always second-order and ghost-free, which is an important feature for quantum investigations. This fact allows us to explore in detail the equations governing the evolution of tensorial perturbations on the brane. We find that with an appropriate choice of variables and some standard redefinitions, these equations can be put in a form which is formally identical to that found in the case of GR. Their stability is proved by showing the positive definiteness of the effective Hamiltonian associated to the higher-dimensional modes. This result confirms the robustness of the braneworld scenario against tensorial perturbations for arbitrary gravity Lagrangian of the form $F(g_{\alpha\beta},R_{\alpha\beta})$ and for arbitrary scalar field dynamics.

\section{Dynamical content}

We consider a rather general family of theories of gravity based on the metric and an independent connection (metric-affine or Palatini approach \cite{go}), with action

\begin{equation}
S=\frac{1}{2\kappa^2}\int d^D x\sqrt{-g}F(g_{\alpha\beta},R_{\alpha\beta})+S_\phi\ ,
\end{equation}
where $S_\phi$ represents the scalar field sector, $\kappa^2$ is a constant with suitable dimensions, $D$ is the number of space-time dimensions and $g$ is the determinant of the space-time metric $g_{\mu\nu}$. The connection appears through the Ricci tensor, which is defined as  $R_{\beta\nu}={R^\alpha}_{\beta\alpha\nu}$, where ${R^\alpha}_{\beta\mu\nu}=\partial_\mu \Gamma^\alpha_{\nu\beta}-\partial_\nu \Gamma^\alpha_{\mu\beta}+\Gamma^\alpha_{\mu\lambda}\Gamma^\lambda_{\nu\beta}-\Gamma^\alpha_{\nu\lambda}\Gamma^\lambda_{\mu\beta}$. Given that the Riemann tensor is antisymmetric only in its two last indices, it is not {\it a priori} guaranteed that for an arbitrary connection $\Gamma^\alpha_{\nu\beta}$ the Ricci tensor has any specific symmetry. For this reason one must bear in mind that there might be an antisymmetric part in $R_{\beta\nu}$. Variation of the above action leads to

\begin{equation} \label{eq:var0}
\delta S= \frac{1}{2\kappa^2}\int d^D x\sqrt{-g}\left[\left(\frac{\partial F}{\partial g^{\mu\nu}}-\frac{1}{2}g_{\mu\nu}F\right)\delta g^{\mu\nu}\nonumber\right. + \left. \frac{\partial F}{\partial R_{\alpha\beta}}\delta R_{\alpha\beta}\right] +\delta S_\phi\ ,
\end{equation}
where the variation of the Ricci tensor is given by

\begin{equation}\label{eq:Riccivar}
\delta R_{\beta\nu}=\nabla_\lambda \delta\Gamma^\lambda_{\nu\beta}-\nabla_\nu \delta\Gamma^\lambda_{\lambda\beta}+2S^\lambda_{\rho\nu}\delta \Gamma^\rho_{\lambda\beta} \
\end{equation}
and $S^\lambda_{\rho\nu}\equiv (\Gamma^\lambda_{\rho\nu}-\Gamma^\lambda_{\nu\rho})/2$ is the torsion tensor. In our metric-affine scenario, the field equations follow from equating to zero the independent variations with respect to the metric and the connection. Given that $\delta R_{\alpha\beta}$ only depends on the connection variation, the metric field equation can be readily extracted from (\ref{eq:var0}), which yields

\begin{equation}
\frac{\partial F}{\partial g^{\mu\nu}}-\frac{1}{2}g_{\mu\nu}F=\kappa^2 T^{(\phi)}_{\mu\nu} \ ,
\end{equation}
where the scalar field  stress-energy tensor $T^{(\phi)}_{\mu\nu}=-\frac{2}{\sqrt{-g}} \frac{\delta S_{\phi}}{\delta g^{\mu\nu}}$ has been included. The connection equation is obtained by inserting (\ref{eq:Riccivar}) into (\ref{eq:var0}) and integrating by parts the first two terms. After elementary manipulations, we find

\begin{equation}
\nabla_\lambda\left(\sqrt{-g}W^{\beta\nu}\right)=2\sqrt{-g}\Bigl[S^\sigma_{\sigma\lambda}W^{\beta\nu}+S^\nu_{\lambda\alpha}W^{\beta\alpha} -\frac{1}{(D-1)}\delta^\nu_\lambda S^\sigma_{\sigma\rho}W^{\beta\rho}\Bigr] ,
\end{equation}
where $W^{\beta\nu}\equiv \partial F/\partial R_{\beta\nu}$.
At this point one must decide whether to keep the torsion and continue with the analysis of the field equations in full generality or introduce some simplification to consider specific cases of interest. In this sense, the analogy with condensed matter systems tells us that torsion is the representation in the continuum  limit of line-like defects such as dislocations and disclinations \cite{Kleinert,Classics}. If torsion is kept in the field equations,  the resulting branes could be seen as entities engendering two or more extra dimensions able to accommodate line-like defects. Since we are focusing on a single extra dimension \cite{rs,gw}, we find it appropriate to deal only with the simplest kind of defects, point-like, which are associated with the non-metricity properties of the connection. We leave the consideration of torsional effects for the future, where a natural direction should concern braneworld scenarios endowed with two extra dimensions \cite{2,2a}. In the case of a single extra dimension, the above equation becomes

\begin{equation}\label{eq:conn}
\nabla_\lambda\left(\sqrt{-g}W^{\beta\nu}\right)=0 .
\end{equation}
A particularly interesting solution of this equation concerns $W^{\beta\nu}$ being a symmetric tensor, which implies that $R_{\beta\nu}=R_{\nu\beta}$. In that case, as shown below, the connection can be immediately solved as the Christoffel symbols of an auxiliary metric which is related to the space-time metric $g_{\mu\nu}$ through a transformation that depends on the stress-energy tensor of the scalar field. To see this explicitly, consider that the Lagrangian $F(g_{\alpha\beta},R_{\alpha\beta})$ is written in terms of the matrix ${M^\mu}_\nu\equiv g^{\mu\lambda}R_{\lambda\nu}$ and traces of its powers. This type of Lagrangians describe the case of $f(R)$ theories \cite{fRor}, where $R$ is the trace of $\hat M$, $f(R,R_{\mu\nu}R^{\mu\nu})$ theories \cite{frqor}, where $R_{\mu\nu}R^{\mu\nu}$ is the trace of $\hat M^2$, Born-Infeld like models \cite{BIor,oo}, and other theories considered in the literature. Using the matrix $\hat M$, we find that
\begin{eqnarray}\label{eq:FM1}
\frac{\partial F}{\partial g^{\mu\nu}} &=& \frac{1}{2}\left[{(F_M)_\mu}^\lambda R_{\nu\lambda}+{(F_M)_\nu}^\lambda R_{\mu\lambda} \right]\\
\frac{\partial F}{\partial R_{\mu\nu}} &=& g^{\mu\lambda} {(F_M)_\lambda}^\nu \ , \label{eq:FM2}
\end{eqnarray}
where ${(F_M)_\lambda}^\nu\equiv \frac{\partial F}{\partial {M^\lambda}_\nu}$. In some cases, the symmetry of $R_{\mu\nu}$ follows directly from the torsionless condition (see \cite{Olmo:2013lta}), though in general it should be seen as an additional simplification, which we assume from now on.  Using (\ref{eq:FM1}) and the definition of $\hat M$ to replace $R_{\mu\nu}$ by $g_{\mu\lambda}{M^\lambda}_\nu$, the metric
field equations can be written as
\begin{equation}\label{eq:EOM-M}
\frac{1}{2}\Big[{(F_M)_\mu}^\lambda g_{\nu\kappa}{M^\kappa}_\lambda +{(F_M)_\nu}^\lambda g_{\mu\kappa}{M^\kappa}_\lambda \Big] -\frac{1}{2}g_{\mu\nu}F =\kappa^2 T^{(\phi)}_{\mu\nu} \ .
\end{equation}
These equations allow to obtain $\hat M$ as an algebraic function of $T^{(\phi)}_{\mu\nu}$ and $g_{\mu\nu}$. As a result, the term $\sqrt{-g}W^{\beta\nu}$ in (\ref{eq:conn}) does not depend explicitly on the connection, which allows to find a solution for $\Gamma^\alpha_{\mu\nu}$ using elementary algebraic manipulations. In fact, the identification

\begin{equation}
\sqrt{-g}g^{\mu\lambda} {(F_M)_\lambda}^\nu=\sqrt{-q}q^{\mu\nu} \ ,
\end{equation}
leads to

\begin{equation}
q^{\mu\nu}=\frac{1}{|\hat F_M|^{\frac{1}{D-2}}}g^{\mu\lambda} {(F_M)_\lambda}^\nu \label{eq:q-g} ; q_{\mu\nu}=|\hat F_M|^{\frac{1}{D-2}}{(F^{-1}_M)_\mu}^\lambda g_{\lambda\nu}
\end{equation}
where $|\hat F_M|$ represents the determinant of the (invertible) matrix ${(F_M)_\lambda}^\nu $. This turns the connection equation (\ref{eq:conn}) into the well-known form

\begin{equation}\label{eq:LC}
\nabla_\lambda \left[\sqrt{-q}q^{\beta\nu}\right]=0 \ ,
\end{equation}
which implies that the solution for $\Gamma^\alpha_{\mu\nu}$ is given by the Christoffel symbols of the auxiliary metric $q_{\alpha\beta}$. This result also means that $R_{\mu\nu}=R_{\mu\nu}(q)$. Note that via Eq.(\ref{eq:q-g}) the relation between $g_{\mu\nu}$ and $q_{\mu\nu}$ is purely algebraic and depends only on the scalar field stress-energy tensor.

When $T^{(\phi)}_{\mu\nu}$ is diagonal, the metric $g_{\mu\nu}$ can be written in diagonal form and (\ref{eq:EOM-M}) suggests that $\hat M$ and $\hat F_M$  also inherit that structure. One can then write (\ref{eq:EOM-M}) in the more compact form

\begin{equation}\label{eq:EOM-M2}
{(F_M)_\mu}^\lambda R_{\nu\lambda}=\kappa^2\left(\frac{F}{2\kappa^2}g_{\mu\nu}+ T^{(\phi)}_{\mu\nu} \right) \ .
\end{equation}
By raising one index on this equation with $g^{\alpha\mu}$ and using (\ref{eq:q-g}), we finally get

\begin{equation}\label{eq:Rmn-q}
{R_\nu}^\alpha(q)=\frac{\kappa^2}{|\hat F_M|^{\frac{1}{D-2}}}\left(\LL_G{\delta_{\nu}}^\alpha+ {T^{(\phi) \alpha}_\nu} \right) \ ,
\end{equation}
where $\LL_G$ represents the gravity Lagrangian of the theory under consideration. This Einstein-like representation of the field equations is valid for GR, $f(R)$, and many other theories \cite{fRor,frqor,BIor,oo}. It puts forward that $q_{\alpha\beta}$ satisfies second-order equations and, given the algebraic relations (\ref{eq:q-g}), it follows that the dynamics of $g_{\alpha\beta}$ is also second-order. When $ {T^{(\phi) \alpha}_\nu}=0$, (\ref{eq:Rmn-q}) recovers Einstein's equations in vacuum (with possibly a cosmological constant term), which are clearly ghost-free. This last point can be verified by setting $T^{(\phi)}_{\mu\nu}$ to zero, raising one index in Eq.(\ref{eq:EOM-M}) with the metric $g^{\alpha\mu}$, and computing traces of powers of the resulting object. Those traces represent algebraic equations that relate the scalars $R\equiv {M^\lambda}_\lambda$, $R_{\mu\nu}R^{\mu\nu}\equiv {[M^2]^\lambda}_\lambda$, and so on. With a sufficient number of equations, which depends on the space-time dimension, one concludes that all such scalars in vacuum must be constants. As a result, the matrix ${M^\alpha}_\beta$ must be proportional to the identity matrix, making in this way $q_{\mu\nu}=\alpha g_{\mu\nu}$, with $\alpha$ being an (irrelevant) constant. One then readily verifies that Eq. (\ref{eq:Rmn-q}) coincides with Einstein's equations in vacuum.

\section{Background equations}

We now assume the space-time to be $(d+1)$-dimensional, with a single extra dimension, and write the line element for the metric $g_{\mu\nu}$ in the form \cite{rs}

\be \label{eq:metric-g}
ds^2=a^2(y)\eta_{ab} dx^a dx^b + dy^2,
\en
being $\eta_{ab}$ the metric of a $d-$dimensional space of constant curvature $K$.
In a thick brane scenario, one assumes that a scalar field
lives in the extra dimension with most of its energy density confined around the  hypersurface $y=0$, which defines the (thick) brane \cite{gw,f,c}. Here the Lagrangian density in the scalar field action, $S_{\phi}=\int d^{d+1}x \sqrt{-g} \LL$, is supposed to be a function $\LL =\LL(\phi,X)$, with $X\equiv g^{\alpha\beta}\partial_\alpha\phi \partial_\beta\phi$. This form allows that the scalar field engenders generalized kinematics, including the case of k-fields \cite{kfield}. Given that $\phi=\phi(y)$, one  finds that the stress-energy tensor of the scalar field can generically be written as

\be \label{eq:Tmunu}
{T^{(\phi)\nu}_\mu}=
\left(
\begin{array}{cc}
T_+(y)  I_{d \times d} &  \hat{0} \\
\hat{0}  & T_-(y)  \\
\end{array}
\right),
\en
where $\hat{I}_{d \times d}$ is the $d \times d$ identity matrix, $T_+=-\LL(\phi,X)/2$, and $T_-=\LL_X \phi_y^2+T_+$, with the definition $\LL_X \equiv d\LL/dX$. As explained above, the diagonal character of ${T^{(\phi)\nu}_\mu}$ induces a diagonal ${(F_M)_\mu}^\lambda$, which allows us to define

\begin{eqnarray} \label{eq:Omunu}
{\Omega_\mu}^\lambda\equiv |\hat F_M|^{\frac{1}{D-2}}{(F^{-1}_M)_\mu}^\lambda&=&
\left(
\begin{array}{cc}
\Omega_+  I_{d \times d} &  \hat{0} \\
\hat{0}  & \Omega_-  \\
\end{array}
\right) \ ,
\end{eqnarray}
with $q_{\mu\nu}={\Omega_\mu}^\lambda g_{\lambda\nu}$, and  $\Omega_\pm$ being functions of $\phi$ and $X$.  We can thus define an auxiliary line element for $q_{\mu\nu}$ of the form
\be \label{eq:metric-q}
ds^2=\tilde{a}^2(y)\eta_{ab} dx^a dx^b + d\tilde{y}^2,
\en
where $\tilde{a}^2(y)=a^2(y) \Omega_+$ and $d\tilde{y}^2=\Omega_-dy^2$ follow immediately from Eq.(\ref{eq:Omunu}). Using this to construct ${R_\nu}^\alpha(q)$, we find that
${R_a}^b\equiv [(d-1)K-(H_{\tilde{y}}+d \cdot H^2)]{\delta_a}^b$ and
${R_{\tilde{y}}}^{\tilde{y}}\equiv -d \cdot (H_{\tilde{y}}+H^2)$, where $H\equiv {\tilde{a}_{\tilde{y}}}/{\tilde{a}}$, and $K$ is the (constant) curvature of the $d$-dimensional brane. Inserting this in (\ref{eq:Rmn-q}), the background equations can thus be written as

\begin{eqnarray}\label{eq:bgd0}
d(d-1)[K-H^2]&=&\frac{\kappa^2}{|\Omega|^{1/2}}\Bigl[(d-1)\LL_G+d \cdot T_+-T_-\Bigr] \;\;\;\;\;\\
(d-1)[K+H_{\tilde{y}}]&=& \frac{\kappa^2}{|\Omega|^{1/2}}\left(T_+-T_-\right) \ .
\end{eqnarray}
A first-order formulation is possible for the case $K=0$ if we define a superpotential $W(\phi)$ such that $H=-W(\phi)/(d-1)$ \cite{bglm}. The case $K\neq0$ is more complicated and we put it aside, recalling that it could perhaps follow the lines of \cite{fab}. For $K=0$ the above equations become

\begin{equation}\label{eq:bgd1}
\LL_X \phi_y=\frac{\Omega_+^{d/2}}{\kappa^2}W_\phi \frac{d}{d-1} W^2(\phi) = \frac{\kappa^2}{|\Omega|^{1/2}}\Bigl(\LL_X \phi_y^2 +(d-1)\Bigl[\frac{\LL(\phi,X)}{2}-\LL_G\Bigr]\Bigr) \ .
\end{equation}
It is easy to see that for the Einstein-Hilbert Lagrangian, $\LL_G=R/2\kappa^2$, the above equations recover the well-known results of GR (in $D=d+1=5$). In this case, we get $\Omega_\pm=1$ and $\LL_G=-(\LL_X \phi_y^2-5\LL/2)/3$, which lead to $\LL_X \phi_y=W_\phi/\kappa^2$ and $\frac{4}{3}W^2(\phi)=\kappa^2(2\LL_X\phi_y^2-\LL)$. For the particular case of a canonical scalar field, $\LL=X+2V$, from this last relation we get $2\kappa^2V(\phi)=\frac{1}{\kappa^2}W_\phi^2-\frac{4}{3}W^2$, which is in agreement with the results presented in \cite{f,c}. In the more general case in which the scalar field Lagrangian can be split into a kinetic part plus a potential term, the first of the above equations can, in principle, be used to isolate $V(\phi)$ as a function of $\phi_y$ and $W_\phi$. Inserting the result in the second, we find an (implicit) first-order equation for $\phi_y$ as a function of $W(\phi)$ and $W_\phi$. The fact that the scalar field is generically governed by a first-order equation suggests that supersymmetric extensions could exist for this type of theories.

\section{Tensorial perturbations}

Using Gaussian normal coordinates, the effect of tensorial perturbations on the line element (\ref{eq:metric-g}) can be parameterized as

\begin{equation}\label{eq:ds2}
ds^2=a^2(y)\left(\eta_{ab}+h_{ab} \right)dx^adx^b+dy^2 \ ,
\end{equation}
where $\delta g_{ab}=a^2(y) h_{ab}$ and $\delta g_{ay}=0=\delta g_{yy}$ account for the tensorial perturbations.
For the auxiliary metric, we must thus have

\begin{equation}\label{eq:dst2}
d\tilde{s}^2=\tilde{a}^2(\tilde{y})\left(\eta_{ab}+h_{ab} \right)dx^adx^b+d\tilde{y}^2 \ .
\end{equation}
Since the tensorial perturbations are restricted to the brane modes, any contraction of the form $\partial_\mu \phi \delta g^{\mu\nu}$ will be zero, as can be easily understood from the structure of (\ref{eq:ds2}) and the fact that only $\partial_y \phi$ is nonzero. As a result, perturbation of (\ref{eq:Rmn-q}) leads simply to

\begin{equation}
\delta {R_\mu}^\nu (q)=0 \ \leftrightarrow \ \delta R_{\mu\nu}(q)={R_\mu}^\beta t_{\beta\nu} \ ,
\end{equation}
where $t_{ab}=\tilde{a}^2 h_{ab}$ is the only nonzero component of $t_{\beta\nu}$. Using standard covariant perturbation methods, one finds that

\begin{eqnarray}
\delta R_{\mu\nu}(q)&\equiv& -\frac{1}{2}q^{\alpha\beta} \nabla_\alpha \nabla_\beta t_{\mu\nu} + \frac{1}{2}\left(\nabla_\mu\nabla_\lambda \hat t^\lambda_\nu+\nabla_\nu\nabla_\lambda \hat t^\lambda_\mu\right) \\
&+&\frac{1}{2}\left(R_{\mu\alpha}t^\alpha_\nu+R_{\nu\alpha}t^\alpha_\mu\right)-{R^\alpha}_{\mu\beta\nu}t^\beta_\alpha \nonumber
\end{eqnarray}
where $\hat t^\lambda_\mu\equiv t^\lambda_\mu-\frac{1}{2}t^\alpha_\alpha$. As usual, we will adopt a traceless and transverse gauge, where $t^\alpha_\alpha=0$ and $\nabla_\lambda t^\lambda_\mu=0$. The transverse gauge can also be written as $\nabla_a t^a_b=0$, with $\nabla_a$ referred to the metric $q_{ab}$, and as $\nabla_i h^i_j=0$, with $\nabla_i$ referred to the metric $\eta_{ij}$ and $h^i_j=h_{ik}\eta^{kj}$.

From the diagonal structure of (\ref{eq:Tmunu}) and Eq.~(\ref{eq:Rmn-q}), it follows that ${R_\mu}^\nu(q)$ can be written in a diagonal form analogous to (\ref{eq:Tmunu}) and  (\ref{eq:Omunu}). Together with the fact that $t^\mu_\nu$ only has non-zero contributions in the brane sector, one can easily verify that $\frac{1}{2}\left(R_{\mu\alpha}t^\alpha_\nu+R_{\nu\alpha}t^\alpha_\mu\right)={R_\mu}^\beta t_{\beta\nu}$. As a result, the equations governing the tensorial perturbations boil down to (we note that similar manipulations can be used to study tensorial peturbations in cosmological backgrounds \cite{Jimenez:2015caa})
\begin{equation}\label{eq:tenpert0}
 \frac{1}{2}q^{\alpha\beta} \nabla_\alpha \nabla_\beta t_\mu^\nu+q^{\nu\lambda}{R^\alpha}_{\mu\beta\lambda}t^\beta_\alpha=0 \ .
\end{equation}
Remarkably, this result is valid for any ${T^{(\phi)\nu}_\mu}$ of the form (\ref{eq:Tmunu}).
With elementary (though lengthy) calculations, we find
\begin{eqnarray}
^{(d+1)}{R^m}_{anb}&=& ^{(d)}{R^m}_{anb}+H^2(\delta^m_b q_{an}-\delta^m_n q_{ab}) \\
q^{\tilde{y}\tilde{y}}\nabla_{\tilde{y}}\nabla_{\tilde{y}} t_a^b&=& \partial_{\tilde{y}\tilde{y}}t_a^b \\
q^{mn}\nabla_m\nabla_n t_a^b&=& ^{(d)}\Box t_a^b+d H\partial_{\tilde{y}}t_a^b-2H^2 t_a^b \ ,
\end{eqnarray}
where $H\equiv  \tilde{a}_{\tilde{y}}/\tilde{a}$. With these results, (\ref{eq:tenpert0}) becomes
\begin{equation}\label{eq:tpy}
\partial_{\tilde{y}\tilde{y}}t_a^b+d H\partial_{\tilde{y}}t_a^b+^{(d)}\Box t_a^b+ {^{(d)}}{R^m}_{anc}q^{cb}t_m^n=0 \ .
\end{equation}
It is customary in the literature to model the brane as a maximally symmetric space of constant curvature $K=\pm1,0$, in analogy with the properties of our universe at large scales. Following this idealization,  we get $^{(d)}{R^m}_{anc}=-K(\delta^m_c \eta_{an}-\delta^m_n \eta_{ac})$, which turns the last term of the above equation into $-2K t_a^b/\tilde{a}^2$. It is also common in the literature to introduce a new coordinate such that the line element (\ref{eq:dst2}) becomes
 \begin{equation}\label{eq:dst2z}
d\tilde{s}^2=\tilde{a}^2\left[\left(\eta_{ab}+h_{ab} \right)dx^adx^b+dz^2\right] \ .
\end{equation}
By doing this, (\ref{eq:tpy}) takes the final form
\begin{equation}\label{eq:tpya}
^{(\eta)}\Box {h_a}^b+\partial_{zz}{h_a}^b+(d-1)\Hcal\partial_{z}{h_a}^b-2K{h_a}^b=0 \ ,
\end{equation}
where $^{(\eta)}\Box $ is computed using the $\eta_{ij}$ metric and $\Hcal\equiv \tilde{a}_z/\tilde{a}$. This last equation is formally identical to that given, for instance, in \cite{Kobayashi:2001jd} for the case of GR in $4+1$ dimensions. Assuming that ${h_a}^{b}=X(z){\epsilon_a}^b(t,\vec{x})$,  (\ref{eq:tpya}) splits in two equations of the form
\begin{eqnarray}
X_{zz}+3\Hcal X_z+p^2 X&=&0 \label{eq:p1} \label{eq:Xz}\\
^{(\eta)}\Box {\epsilon_a}^b-2K{\epsilon_a}^b-p^2{\epsilon_a}^b&=&0 \ , \label{eq:epsilon}
\end{eqnarray}
where $p^2$ is a constant. At this point, it is important to note that the ${\epsilon_a}^b(t,\vec{x})$ part of the tensorial modes satisfies an equation which is identical to that found in GR. This equation only depends on the brane coordinates ($t,\vec{x}$) and the brane metric $\eta_{ab}$ but is insensitive to the gravity and matter Lagrangians. The dependence on the extra dimension and on the details of the gravity and matter models manifests itself through the dependence of the function $X(z)$ on the coordinate $z$, defined via $dz^2=\Omega_- dy^2/(\Omega_+ a^2(y))$, and on the function $\Hcal\equiv \tilde{a}_z/\tilde{a}$. Focusing now on Eq.(\ref{eq:Xz}) and redefining $X=\tilde{a}^{-\frac{(d-1)}{2}}Y$, we find that Eq.(\ref{eq:p1}) can be written as
\begin{equation}\label{eq:Eff_Sch}
-Y_{zz}+V_{eff}(z)Y=p^2Y \ ,
\end{equation}
where

\begin{equation}
V_{eff}=\frac{(d-1)}{2}\Hcal_z+\frac{(d-1)^2}{4}\Hcal^2.
\end{equation}
It is now immediate to see that the differential operator on the left-hand side of (\ref{eq:Eff_Sch}) can be written  as the product $S^\dag S$, where $S=\frac{d}{dz}-\frac{(d-1)}{2}\Hcal$. This decomposition implies that $S^\dag S$ is a non-negative operator, ensuring that $p^2\geq0$ and guaranteeing stability of the gravity sector. The zero mode arises for $p=0$ and can be readily solved using (\ref{eq:Xz}) to obtain
\begin{equation}\label{eq:zeromode}
X(z)=X_0+\int^z\frac{C}{\tilde{a}(z')^3}dz' \ ,
\end{equation}
where $X_0$ and $C$ are integration constants. In order to avoid pathologic behaviors as one moves away from the brane, it is necessary to set $C\to 0$, which turns $X(z)$ into a constant. The constancy of $X(z)$ for $p=0$ shows that the zero mode of tensorial perturbations is the same for all Palatini theories of the form discussed in this work. This puts forward that the stability of the zero modes observed in the case of GR is a generic prediction insensitive to the details of the scalar field and gravity Lagrangian chosen, thus supporting the robustness of braneworld scenarios against tensorial perturbations.

\section{Summary}

In this work we have investigated the background structure and tensorial perturbations for generic braneworld scenarios in $(d+1)$-dimensional geometries with a single extra dimension of infinite extent. The family of theories of gravity that we considered is constructed trading the Ricci scalar $R=g^{\alpha\beta}R_{\alpha\beta}$ with a generic function $F(g_{\alpha\beta}, R_{\alpha\beta})$ and assuming that the metric and affine structures are {\it a priori} independent (metric-affine or Palatini formalism \cite{go}), an aspect motivated by condensed matter systems with defects. Remarkably, this has led to a generic set of ghost-free second-order equations for the background, regardless of the particular form of the gravity Lagrangian.  We have shown that by a suitable choice of variables, the scalar field and background equations can be written as a set of first-order equations for the warp factor,  $H=-W(\phi)/(d-1)$, and the scalar field once a superpotential function $W(\phi)$ is introduced. This property suggests that supersymmetric extensions of these theories might be possible.

We have also shown that the gravity sector in these theories is stable under linear perturbations and develops a zero mode (the graviton) bound to the brane with exactly the same properties as the massless gravitons of GR. Remarkably, these results are robust in the sense that they hold no matter the specific forms of both the gravitational Lagrangian, $\LL_G$, and scalar field model with generalized kinematics, $\LL(\phi,X)$, chosen. The results presented here can be used to study specific models, such as Born-Infeld gravity and its extensions \cite{BIor,oo}, as well as other theories considered in recent investigations \cite{ba}. We shall further report on this elsewhere.

\section*{Acknowledgments}

D.B., L.L. and R.M. would like to thank CAPES and CNPq for financial support. G.J.O. is supported by a Ramon y Cajal contract, the Spanish grant FIS2011-29813-C02-02, the Consolider Program CPANPHY-1205388,  and  the i-LINK0780 grant of the Spanish Research Council (CSIC). D.R.-G. is supported by the NSFC (Chinese agency) grants No. 11305038 and 11450110403, the Shanghai Municipal Education Commission grant for Innovative Programs No. 14ZZ001, the Thousand Young Talents Program, and Fudan University. The authors also acknowledge funding support of CNPq project No. 301137/2014-5.

\end{document}